\documentclass[a4paper,twocolumn,english,8pt,amssymb,nofootinbib,superscriptaddress]{revtex4}
\usepackage{lmodern}
\usepackage{lmodern}
\usepackage[T1]{fontenc}
\usepackage[latin9]{inputenc}
\usepackage{babel}
\usepackage{amsmath}
\usepackage{amssymb}
\usepackage{esint}
\usepackage[unicode=true,pdfusetitle,
 bookmarks=true,bookmarksnumbered=false,bookmarksopen=false,
 breaklinks=false,pdfborder={0 0 1},backref=false,colorlinks=false]
 {hyperref}

\makeatletter

\pdfpageheight\paperheight
\pdfpagewidth\paperwidth

\@ifundefined{textcolor}{}
{%
 \definecolor{BLACK}{gray}{0}
 \definecolor{WHITE}{gray}{1}
 \definecolor{RED}{rgb}{1,0,0}
 \definecolor{GREEN}{rgb}{0,1,0}
 \definecolor{BLUE}{rgb}{0,0,1}
 \definecolor{CYAN}{cmyk}{1,0,0,0}
 \definecolor{MAGENTA}{cmyk}{0,1,0,0}
 \definecolor{YELLOW}{cmyk}{0,0,1,0}
}

\usepackage{babel}

\usepackage{latexsym}\usepackage{bm}

\makeatother

\begin{document}
\title{Comment on ``Einstein-Gauss-Bonnet Gravity in 4-Dimensional Space-Time'' }
\author{Metin Gürses}
\email{gurses@fen.bilkent.edu.tr}

\affiliation{{\small{}Department of Mathematics, Faculty of Sciences}\\
{\small{}Bilkent University, 06800 Ankara, Turkey}}
\author{Tahsin Ça\u{g}r\i{} \c{S}i\c{s}man}
\email{tahsin.c.sisman@gmail.com}

\affiliation{Department of Astronautical Engineering,\\
 University of Turkish Aeronautical Association, 06790 Ankara, Turkey}
\author{Bayram Tekin}
\email{btekin@metu.edu.tr}

\affiliation{Department of Physics, Middle East Technical University, 06800 Ankara,
Turkey}

\maketitle
In a recent Letter \citep{GlavanLin}, a general covariant four-dimensional
modified gravity that propagates only a massless spin-2 graviton and
bypasses Lovelock's theorem \citep{Lovelock1} was claimed to exist.
Here we show that this claim is not correct. The suggested theory
is a limit of the Einstein-Gauss-Bonnet theory with the field equations
\[
\lim_{D\rightarrow4}\left[\frac{1}{\kappa}\left(R_{\mu\nu}-\frac{1}{2}g_{\mu\nu}R+\Lambda_{0}g_{\mu\nu}\right)+\frac{\alpha}{D-4}{\cal H}_{\mu\nu}\right]=0,
\]
where the ``Gauss-Bonnet (GB) tensor'' (which vanishes identically
in four dimensions ) reads \citep{DeserTekinPRD} 
\begin{eqnarray}
{\cal H}_{\mu\nu}= & 2\Bigg[RR_{\mu\nu}-2R_{\mu\alpha\nu\beta}R^{\alpha\beta}+R_{\mu\alpha\beta\sigma}R_{\nu}^{\phantom{\nu}\alpha\beta\sigma}-2R_{\mu\alpha}R_{\nu}^{\alpha}\nonumber \\
 & -\frac{1}{4}g_{\mu\nu}\left(R_{\alpha\beta\rho\sigma}R^{\alpha\beta\rho\sigma}-4R_{\alpha\beta}R^{\alpha\beta}+R^{2}\right)\Bigg].\label{eq:EoM}
\end{eqnarray}
For the $D\rightarrow4$ limit to work even at the formal level, there
must exist a new tensor $\mathcal{Y}_{\mu\nu}$ such that one has
\begin{equation}
{\cal H}_{\mu\nu}=\left(D-4\right)\mathcal{Y}_{\mu\nu},\label{eq:Y}
\end{equation}
and as $D\rightarrow4$, this new tensor should not vanish and should
have a smooth limit. One can show that \citep{Gurses} ${\cal H}_{\mu\nu}$
decomposes as 
\begin{equation}
\frac{{\cal H}_{\mu\nu}}{D-4}=2\frac{{\cal {L}_{\mu\nu}}}{D-4}+\frac{2\left(D-3\right)}{\left(D-1\right)\left(D-2\right)}S_{\mu\nu},\label{onemli}
\end{equation}
where ${\cal {L}_{\mu\nu}}=C_{\mu\alpha\beta\gamma}\,C_{\nu}\,^{\alpha\beta\gamma}-\frac{1}{4}C_{\alpha\beta\gamma\delta}\,C^{\alpha\beta\gamma\delta}\,g_{\mu\nu}$
and 
\begin{align*}
 & S_{\mu\nu}=-\frac{2\left(D-1\right)}{\left(D-3\right)}C_{\mu\rho\nu\sigma}R^{\rho\sigma}-\frac{2\left(D-1\right)}{\left(D-2\right)}R_{\mu\rho}R_{\nu}^{\rho}\\
 & +\frac{D}{\left(D-2\right)}R_{\mu\nu}R+\frac{\left(D-1\right)}{\left(D-2\right)}g_{\mu\nu}\left(R_{\rho\sigma}R^{\rho\sigma}-\frac{D+2}{4\left(D-1\right)}R^{2}\right).
\end{align*}
The $S_{\mu\nu}$ part in Eq.~(\ref{onemli}) is smooth in the $D\rightarrow4$
limit, but the ${\cal {L}_{\mu\nu}}$ part is undefined ($0/0$) and
discontinuous in the sense that ${\cal {L}_{\mu\nu}}$ is \textit{identically}
zero in four dimensions and nontrivial above four dimensions. If one
naively takes the limit by dropping the ${\cal {L}_{\mu\nu}}$ part
in Eq.~(\ref{onemli}), then one loses the Bianchi identity since
$\nabla_{\mu}S^{\mu\nu}\ne0$, which is not acceptable if gravity
is expected to couple to a conserved source.

The easiest way \citep{Gurses} to see that such a tensor $\mathcal{Y}_{\mu\nu}$
does not exist in four dimensions is to employ the first order form
of the theory. The GB part of the action (without any factors) 
\begin{align*}
I_{{\rm GB}} & =\int_{{\mathcal{M}}_{D}}\epsilon_{a_{1}a_{2}...a_{D}}R^{a_{1}a_{2}}\wedge R^{a_{3}a_{4}}\wedge e^{a_{5}}\wedge e^{a_{6}}...\wedge e^{a_{D}}
\end{align*}
when varied with respect to the vielbein yields zero in $D=4$ dimensions;
and the following $D-1$ form in $D>4$, 
\[
\mathcal{E}_{a_{D}}=(D-4)\epsilon_{a_{1}a_{2}...a_{D}}R^{a_{1}a_{2}}\wedge R^{a_{3}a_{4}}\wedge e^{a_{5}}\wedge e^{a_{6}}...\wedge e^{a_{D-1}}.
\]
To relate this to ${\cal H}_{\mu\nu}$ (\ref{eq:EoM}), one can recast
the last expression in spacetime indices and take its Hodge dual to
get a 1-form 
\begin{eqnarray}
*{\mathcal{E}}_{\nu}= & \frac{\left(D-4\right)}{4}\epsilon_{\mu_{1}\mu_{2}...\mu_{D-1}\nu}\epsilon^{\sigma_{1}...\sigma_{4}\mu_{5}...\mu_{D-1}}\,_{\mu_{D}}\nonumber \\
 & R_{\phantom{\mu_{1}\mu_{2}}\sigma_{1}\sigma_{2}}^{\mu_{1}\mu_{2}}R_{\phantom{\mu_{3}\mu_{4}}\sigma_{3}\sigma_{4}}^{\mu_{3}\mu_{4}}dx^{\mu_{D}},
\end{eqnarray}
from which one defines the rank-2 tensor ${\mathcal{E}}_{\nu\alpha}$
as $*{\mathcal{E}}_{\nu}=:{\mathcal{E}}_{\nu\alpha}dx^{\alpha}$ whose
explicit form is 
\begin{eqnarray}
{\mathcal{E}}_{\nu\alpha}= & \frac{\left(D-4\right)}{4}\epsilon_{\mu_{1}\mu_{2}...\mu_{D-1}\nu}\epsilon^{\sigma_{1}...\sigma_{4}\mu_{5}...\mu_{D-1}}\,_{\alpha}\nonumber \\
 & R_{\phantom{\mu_{1}\mu_{2}}\sigma_{1}\sigma_{2}}^{\mu_{1}\mu_{2}}R_{\phantom{\mu_{3}\mu_{4}}\sigma_{3}\sigma_{4}}^{\mu_{3}\mu_{4}}.
\end{eqnarray}
It is clear from this expression that a $(D-4)$ factor arises only
in $D>4$ dimensions; namely, even if the front factor can be canceled
by multiplying with a $1/(D-4)$ as suggested in Ref.~\citep{GlavanLin},
the $\epsilon$ tensors in the expression explicitly show the dimensionality
of the spacetime to be $D>4$. If one tries to get rid of the epsilon
tensors, then one loses the front factor. In fact, expressing the
epsilon factors in terms of generalized Kronecker delta tensors, one
arrives at ${\mathcal{E}}_{\nu\alpha}=2\left(D-4\right)!{\mathcal{H}}_{\nu\alpha}$,
in which the front factor transmutes to $\left(D-4\right)$ factorial.
So the upshot is that there is no nontrivial $\mathcal{Y}_{\mu\nu}$
(\ref{eq:Y}) in four dimensions as is required for the claim of Ref.~\citep{GlavanLin}
to work. Our result is consistent with the Lovelock's theorem which
rigorously shows that in four dimensions, the only second rank symmetric,
covariantly conserved tensor that is at most second order in derivatives
of the metric tensor (and this is required for a massless graviton
and no other degrees of freedom), besides the metric, is the Einstein
tensor. Therefore a simple rescaling of the coefficient in the EGB
theory as was suggested in Ref.~\citep{GlavanLin} does not yield
covariant equations of a massless spin-2 theory in four dimensions.
The abovementioned lack of continuity of the EGB theory at $D=4$
can also be seen from various complementary analyses \citep{Pang,Aoki}
which try to obtain a well-defined limit and end up with an extra
scalar degree of freedom besides the massless graviton. The resulting
theory depends on how one defines the limit supporting our arguments
here.

\end{document}